\documentclass{elsart3}
\usepackage{graphicx}
\usepackage{amsmath}

\begin{document}

\begin{frontmatter}

\title{Effects of two-site composite excitations in the Hubbard model}
\author{A. Avella}
\author{F. Mancini}
\author{S. Odashima\corauthref{}}
\ead{odashima@sa.infn.it}
\ead[url]{http://www.scs.sa.infn.it}
\corauth[]{Corresponding author. Tel: +39 089 965228; Fax: +39 089 965275}
\address{Dipartimento di Fisica ``E.R. Caianiello'' - Unit\`a INFM di Salerno,
Universit\`a degli Studi di Salerno, I-84081 Baronissi
(SA), Italy} 

\begin{abstract}
The electronic states of the Hubbard model are investigated by use
of the Composite Operator Method. In addition to the Hubbard
operators, two other operators related with two-site composite
excitations are included in the basis. Within the present
formulation, higher-order composite excitations are reduced to the
chosen operatorial basis by means of a procedure preserving the
particle-hole symmetry. The positive comparison with numerical
simulations for the double occupancy indicates that such
approximation improves over the two-pole approximation.
\end{abstract}

\end{frontmatter}

The analysis of highly correlated electron systems
still reports many unsolved issues in spite of the large number of
efforts which have been made for several decades. In general,
difficulties come from the treatment of the collective excitations
emerging in these systems, especially near the Mott-Hubbard
transition. In this paper, we study the electronic states of the
Hubbard model by use of the Composite Operator Method
\cite{COM01,COM02,COM03}, which has shown to be capable to describe the
physics of strongly correlated systems.
%
Within projection techniques, going beyond the two-pole approximation, which
implies the use of a local basis, is a very hard task. As first step along this direction,
we here first used a four component operatorial basis with two non-local fields
extending over two-sites.

The $d$-dimensional Hubbard Hamiltonian reads as follows,
\[
H = \sum_{ij\sigma}(t_{ij}-\mu\delta_{ij})c^{\dagger}_{\sigma}(i)c_{\sigma}(j)
  +U\sum_{i}n_{\uparrow}(i)n_{\downarrow}(i),
\]
where $c^{\dagger}_{\sigma}(i)$ and $c_{\sigma}(i)$ are creation
and annihilation operators of electrons with spin $\sigma$ at the
site $i$, respectively. $n_{\sigma}(i) = c^{\dagger}_{\sigma}(i)
c_{\sigma}(i)$, $\mu$ is the chemical potential, $t_{ij}=- 2 d t
\alpha_{ij}$, $\alpha[{\bf k}]=\mathcal{F} [\alpha_{ij} ] =\frac{1}{d}
\sum^{d}_{i=1} \cos(k_i a)$, $a$ is the lattice constant,
$\mathcal{F}$ is the Fourier transform, $U$ is the on-site Coulomb
repulsion. We define the following operatorial basis,
\begin{equation}
\psi_{\sigma}(i)=\left( \begin{array}{l}
 \xi_{\sigma}(i) \\
 \eta_{\sigma}(i) \\
 \xi_{s\sigma}(i) \\
 \eta_{s\sigma}(i)
\end{array}\right),
\end{equation}
where $\xi_{\sigma}(i)=c_{\sigma}(i)\left( 1-n_{-\sigma}(i) \right)$ and
$\eta_{\sigma}(i)=c_{\sigma}(i)n_{-\sigma}(i)$ describe
the transitions $n(i)=0$ $\leftrightarrow$ 1 and 1 $\leftrightarrow$
2, respectively. $\xi_{s\sigma}(i)$ and $\eta_{s\sigma}(i)$ are
\begin{equation}
\begin{array}{l}
\xi_{s\sigma}(i)=-n_{-\sigma}(i)\xi^{\alpha}_{\sigma}(i)
 +c^{\dagger}_{-\sigma}(i)c_{\sigma}(i)\xi^{\alpha}_{-\sigma}(i) \\
 \hspace*{14mm} +c_{\sigma}(i)\eta^{\alpha\dagger}_{-\sigma}(i)c_{-\sigma}(i) \vspace*{1mm} \\
\eta_{s\sigma}(i)=-n_{-\sigma}(i)\eta^{\alpha}_{\sigma}(i)
 +c^{\dagger}_{-\sigma}(i)c_{\sigma}(i)\eta^{\alpha}_{-\sigma}(i) \\
 \hspace*{14mm} +c_{\sigma}(i)\xi^{\alpha\dagger}_{-\sigma}(i)c_{-\sigma}(i)
\end{array}
\end{equation}
with $c^{\alpha}_{\sigma}(i)=\sum_{j}\alpha_{ij}c_{\sigma}(j)$.
These operators describe two-site composite excitations which were
not included in the previous work \cite{COM02,COM03}, and are
eigenoperators of the interaction term of the Hamiltonian,
similarly to $\xi_{\sigma}(i)$ and $\eta_{\sigma}(i)$.

The equations of motion of the basis read as
\[
\begin{array}{l}
i\frac{\partial}{\partial t}\xi_{\sigma}(i)=-\mu \xi_{\sigma}(i)
 -zt \left[ c^{\alpha}_{\sigma}(i) +\xi_{s\sigma}(i)+\eta_{s\sigma}(i) \right]  \vspace*{1mm} \\
i\frac{\partial}{\partial t}\eta_{\sigma}(i)=(-\mu+U)\eta_{\sigma}(i)
 +zt \left[ \xi_{s\sigma}(i) +\eta_{s\sigma}(i) \right] \vspace*{1mm} \\
i\frac{\partial}{\partial t}\xi_{s\sigma}(i)=-\mu\xi_{s\sigma}(i) \\
 \hspace*{7mm}-zt \left[ -\frac{2}{z} \eta_{\sigma}(i) -\xi^{\alpha}_{s\sigma}(i)
 -2\eta^{\alpha}_{s\sigma}(i) +DJ_{\xi}(i) \right] \vspace*{1mm} \\
i\frac{\partial}{\partial t}\eta_{s\sigma}(i)=(-\mu+U)\eta_{s\sigma}(i) \\
 \hspace*{7mm}-zt \left[ -\frac{1}{z}\eta_{\sigma}(i)-\xi^{\alpha}_{s\sigma}(i)+DJ_{\eta}(i) \right].
\end{array}
\]
where $z=2d$ is the coordination number. $DJ_{\xi}$ and
$DJ_{\eta}$ are 3-site irreducible operators. By
\emph{irreducible} we mean that all local and two-site
contributions have been subtracted. Hereafter, we will neglect
$DJ_{\xi}$ and $DJ_{\eta}$. It is worth mentioning that the
particle-hole symmetry of the model is preserved by this
approximation. Then, the thermal retarded Green's function $G({\bf
k}, \omega) = \mathcal{F} \langle \mathcal{R} [ \psi(i) \
\psi^{\dagger}(j) ] \rangle $ can be expressed as,
\begin{equation}
G({\bf k}, \omega)=\sum_{n=1}^{4}\frac{\sigma_n({\bf k})}{\omega-E_{n}({\bf k})},
\end{equation}
where
\begin{equation}
\begin{array}{l}
 E_{1}({\bf k})=-\mu-zt\alpha[{\bf k}] \\
 E_{2}({\bf k})=-\mu-zt\alpha[{\bf k}]+U \\
 E_{3}({\bf k})=-\mu+zt\alpha[{\bf k}]-zJ_U \\
 E_{4}({\bf k})=-\mu+zt\alpha[{\bf k}]+U+zJ_U,
\end{array}
\end{equation}
with $J_U=1/z(\sqrt{4zt^2+(U/2)^2}-U/2)$.

The spectral densities $\sigma_{n}({\bf k})$ contain two correlation functions:
$\Delta=\langle \xi^{\alpha}_{\uparrow}(i)\xi^{\dagger}_{\uparrow}(i) \rangle
 -\langle \eta^{\alpha}_{\uparrow}(i) \eta^{\dagger}_{\uparrow}(i) \rangle $ and
$p=\frac{1}{4} \langle n_{\nu}(i)n^{\alpha}_{\nu}(i) \rangle
 -\langle ( c_{\uparrow}(i)c_{\downarrow}(i))^{\alpha}
 c^{\dagger}_{\downarrow}(i) c^{\dagger}_{\uparrow}(i) \rangle $.
$\Delta$ can be directly computed in terms of the Green's
function. $p$ and $\mu$ are self-consistently evaluated through
the constraint $\langle \xi (i)\eta^{\dagger} (i) \rangle =0$ and
the equation defining the electron number density $\langle n
\rangle$.

To test the reliability of the present approximation, we
calculated the double occupancy $D=\langle
n_{\uparrow}(i)n_{\downarrow}(i) \rangle$, and compared our
results with the numerical data obtained by the Lanczos method for
a 18-site system \cite{LANCZOS}. The agreement is considerably
good over the whole $U/t$ range and the results show a clear
improvement over the ones obtained by the two-pole approximation \cite{COM03}.
The details of the formulation and more extensive comparisons with
numerical simulations will be presented elsewhere.

\begin{figure}
\begin{center}
\includegraphics[width=70mm]{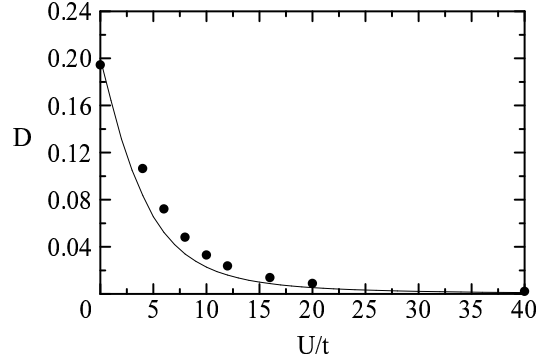}
\end{center}
\caption{The double occupancy $D$ is reported as a function of
$U/t$. $\langle n \rangle =8/9$ and the temperature $T=0$. Full
circles represent the results of Ref.~\cite{LANCZOS}.}
\end{figure}

\end{document}